\documentstyle[aps,prd,preprint,floats,epsfig]{revtex} 
 
\tightenlines    

\begin{document}

\def\bc{{_{\cal C}}}
\def\bu{{_{\cal U}}}
\def\sex{S_{\rm ex}}
\def\sexKst{\sex^{\kstar +}}
\def\sexKstm{\sex^{\kstar -}}
\def\ssign{S_{\rm sign}}
\def\spi{S_\pi}

\def\bbartocbar{\overline{b}\rightarrow\overline{c}u\overline{s}}
\def\bbartoubar{\overline{b}\rightarrow\overline{u}c\overline{s}}
\def\btoc{{b}\rightarrow{c}\overline{u}{s}}
\def\btou{{b}\rightarrow{u}\overline{c}{s}}
\def\bptodzbkp{B^+\rightarrow\overline{D^0}K^+}
\def\bptodzkp{B^+\rightarrow D^0 K^+}
\def\dkpi{D K \pi}
\def\btodkpi{$ B \rightarrow \dkpi $}
\def\bpmtodkpi{$ B^\pm \rightarrow \dzdzbkpmpiz $}
\def\btodkpiall{$ B \rightarrow D^{(\ast)} K^{(\ast)} \pi(\rho) $}
\def\btodkpmpiz{$ B^\pm \rightarrow D^0 K^\pm \pi^0 $}
\def\dcp{D_{1,2}^0}
\def\dcpp{D_{1}^0}
\def\dcpm{D_{2}^0}
\def\dzkpmpiz{D^0 K^\pm \pi^0}
\def\dkpmpiz{D K^\pm \pi^0}
\def\dzkpiz{D^0 K^- \pi^0}
\def\zdkpi{D^+ (K \pi)^-}
\def\dzdzbkpmpiz{D^0(\overline{D^0}) K^\pm \pi^0}
\def\dzkppiz{D^0 K^+ \pi^0}
\def\dzbkppiz{\overline{D^0} K^+ \pi^0}
\def\dzkmpiz{D^0 K^- \pi^0}
\def\dzbkmpiz{\overline{D^0} K^- \pi^0}
\def\dzkpmpiz{D^0 K^\pm \pi^0}
\def\dzbkpmpiz{\overline{D^0} K^\pm \pi^0}
\def\dcpkppiz{\dcp K^+ \pi^0}
\def\dcpkmpiz{\dcp K^- \pi^0}
\def\dcpkpmpiz{\dcp K^\pm \pi^0}
\def\bptodz{B^+ \rightarrow \dzkppiz}
\def\bptodzb{B^+ \rightarrow \dzbkppiz}
\def\bmtodz{B^- \rightarrow \dzkmpiz}
\def\bmtodzb{B^- \rightarrow \dzbkmpiz}
\def\bpmtodz{B^\pm \rightarrow \dzkpmpiz}
\def\bpmtodzb{B^\pm \rightarrow \dzbkpmpiz}
\def\bptodcp{B^+ \rightarrow \dcpkppiz}
\def\bmtodcp{B^- \rightarrow \dcpkmpiz}
\def\bpmtodcp{B^\pm \rightarrow \dcpkpmpiz}
\def\fb{fb$^{-1}$}
\def\lumi{400~\fb}
\def\kstar{{K^{\ast}}}
\def\inv{$\surd$}
\def\pinv{$\overline\surd$}
\def\noinv{}


\newcommand{\beqn}{\begin{eqnarray}}
\newcommand{\eeqn}{\end{eqnarray}}

\begin{flushright}
CSUHEP 02-01\\
DAPNIA-02-95 \\
LAL-02-83
\end{flushright}

\begin{center}
{\large \bf
	Measuring the Weak Phase $\gamma$ in Color 
	Allowed $B\rightarrow DK\pi$ Decays
}

\bigskip
{Roy Aleksan}

{\it Centre d'Etudes Nucleaires, Saclay, DAPNIA/SPP, 
	F-91191 Gif-sur-Yvette, CEDEX, France}

\bigskip
{Troels C. Petersen}

{\it LAL Bat 208, Orsay BP 34, 91898 France}

\bigskip
{Abner Soffer}

{\it Department of Physics, 
	Colorado State University, Fort Collins, CO 80523, U.S.A.}

(\today)
\end{center}

%
%
%
%
%

\begin{abstract}
We present a method to measure the weak phase $\gamma$ in the
three-body decay of charged $B^\pm$ mesons to the final states
$\dkpmpiz$.  These decays are mediated by interfering amplitudes which
are color-allowed and hence relatively large. As a result, large CP
violation effects that could be observed with high statistical
significance are possible. In addition, the three-body decay helps
resolve discrete ambiguities that are usually present in measurements
of the weak phase. The experimental implications of conducting these
measurements with three-body decays are discussed, and the sensitivity
of the method is evaluated using a simulation.
\end{abstract}


\section{Introduction}
CP violation is currently the focus of a great deal of attention.
Since the start of operation of the B-factories, the standard model
description of CP violation via the Cabibbo-Kobayashi-Maskawa (CKM)
matrix~\cite{ref:ckm} is being tested with increasing precision.
BaBar~\cite{ref:babar-sin2b} and Belle~\cite{ref:belle-sin2b} have
recently published measurements of the CKM parameter $\sin(2\beta)$,
where $\beta = \arg{\left(- V_{cd} V_{cb}^\ast/ V_{td}
V_{tb}^\ast\right)}$, verifying the CKM mechanism to within the
experimental sensitivity. Although improved measurements of $\sin(2\beta)$ and
$B_s - \overline B_s$ mixing will probe the theory with greater scrutiny 
during the next few years, the measurement of the other angles of the 
unitarity triangle are necessary for a comprehensive study of CP violation.

Important constraints on the theory will be obtained from measurements
of the CKM phase $\gamma = \arg(-V_{ud} V_{ub}^\ast / V_{cd}
V_{cb}^\ast)$.  A promising method for measuring this phase in the $B$
system has been proposed~\cite{ADK:1}.  Although this method involves
color-allowed decays and hence offers favorable rates, it makes use of
$B_s$ mesons, which are not produced at B-Factories operating at the
$\Upsilon$(4S) resonance.  By contrast, the extraction of $\gamma$
using the B$_u$ and B$_d$ system generally involves decays which are
highly suppressed or difficult to reconstruct. In addition,
these methods are generally subject to an
eight-fold ambiguity due to {\it a-priori} unknown strong
phases~\cite{ADK:1,ref:soffer-ambig}. As a result, obtaining
satisfactory sensitivity requires very high statistics and
necessitates the use of as many decay modes and measurement methods as
possible.

One important class of theoretically clean measurements will make use
of decays of the type $B\rightarrow DK$. Gronau and
Wyler~\cite{ref:GW} have proposed to measure $\sin^2\gamma$ in the
interference between the $\bbartocbar$ decay $\bptodzbkp$ and the
color-suppressed $\bbartoubar$ decay $\bptodzkp$.  Interference
between these amplitudes takes place when the $D$ meson is observed as
one of the CP-eigenstates
\begin{equation}
D_{1,2} \equiv {1\over\sqrt{2}} \left(D^0 \pm
\overline{D^0}\right), 
\label{eq:D12}
\end{equation}
which are identified by their decay products, such as $K^+K^-$ or $K_s
\pi^0$. 
Several variations of this method have been
developed\cite{ref:dunietz,ref:resonance,ref:jk,ref:gr}, including addressing
the effects of doubly Cabibbo-suppressed decays of the $D$
meson~\cite{ref:ADS} and mixing and CP-violation in the neutral $D$
meson system~\cite{ref:SS}, as well as insights to be gained from charm
factory measurements~\cite{ref:soffer-charm}.

A serious difficulty with measuring $\gamma$ using $B\rightarrow DK$
is that the $\bbartoubar$ amplitude is expected to be
extremely small. To a large degree, this is due to the
color-suppression associated with the internal spectator diagram
through which this amplitude proceeds. In the factorization model,
color-suppression of the amplitude is parameterized by the
phenomenological ratio $|a_2 /a_1|$. This ratio is measured to be
about $0.25$~\cite{ref:cleo-bsw} by comparing decay modes which depend
only on color-allowed amplitudes with those that depend on both
color-allowed and color-suppressed amplitudes. With this value of
$|a_2 /a_1|$, one expects the amplitude ratio $|{\cal A}(\bptodzkp )
 / {\cal A}(\bptodzbkp )|$ to be only about 0.1. 
The small branching fraction ${\cal B}(\bptodzkp )$ is therefore very
difficult to measure with adequate precision, resulting in a large
statistical error in the measurement of $\sin^2\gamma$. The recent
observation of the color-suppressed decays $B^0 \rightarrow
\overline{D^{(\ast)0}} \pi^0$, $\overline{D^0}\eta$, and
$\overline{D^0}\omega$ by Belle~\cite{ref:belle-color} and
CLEO~\cite{ref:cleo-color} has raised the possibility that $|a_2/a_1|$
may be effectively larger in some modes. However, significant
suppression is still expected for internal spectator diagrams.

This difficulty has led to attempts to address the problems presented
by color suppression. Dunietz~\cite{ref:dunietz} proposed to apply the
method to the decays $B^0 \rightarrow D K^{*0}$, making use of the
fact that the decay $K^{*0} \rightarrow K^+ \pi^-$ tags the flavor of
the $B^0$.  In this mode, both the $\bbartocbar$ and the $\bbartoubar$
amplitudes are color suppressed, and hence of similar magnitudes,
albeit small. Jang and Ko~\cite{ref:jk} and Gronau and
Rosner~\cite{ref:gr} have devised a method in which the small
branching fraction of the color-suppressed decay $\bptodzkp$ does not
have to be measured directly. Rather, it is essentially inferred by
using the larger branching fractions of the decays $B^0\rightarrow D^-
K^+$, $B^0\rightarrow \overline{D^0} K^0$ and $B^0\rightarrow D_{1,2}
K^0$. Quantitative study suggests that the various alternative methods
are roughly as sensitive as the method of Ref.~\cite{ref:GW}, and are
thus useful for increasing statistics and providing consistency
checks~\cite{ref:soffer-ambig}.

\section{Measuring \boldmath $\gamma$ with Color-Allowed \btodkpi\ Decays} 
\label{sec:method-infinite-stat}
In this paper we investigate a way to circumvent the color suppression
penalty by using $B^\pm$ decay modes which could potentially offer
significantly large branching fractions, as well as large CP
asymmetries.  Similar modes involving neutral $B$ decays can also be
used. For example the final state $ D^0(\overline{D^0}) K^\pm \pi^\mp $
can be analyzed with the same technique as described here.
Some other decays (such as $ B^0 \rightarrow D^- K_s \pi^+ $)
need a different treatment and will be discussed elsewhere ~\cite{xxx}.
The particular decays which are considered here are of the type
\btodkpiall. These three body final states may be obtained by popping a
$q\bar{q}$ pair in color allowed decays.  Although modes where one or
more of the three final state particles is a vector can also be used,
for clarity and simplicity only the mode \btodkpmpiz\ is discussed
here.

\begin{figure}[htb]
\vfill
\begin{minipage}{.45\linewidth}
\begin{minipage}[h]{1.0\textwidth}
\begin{center}
\mbox{\epsfig{file=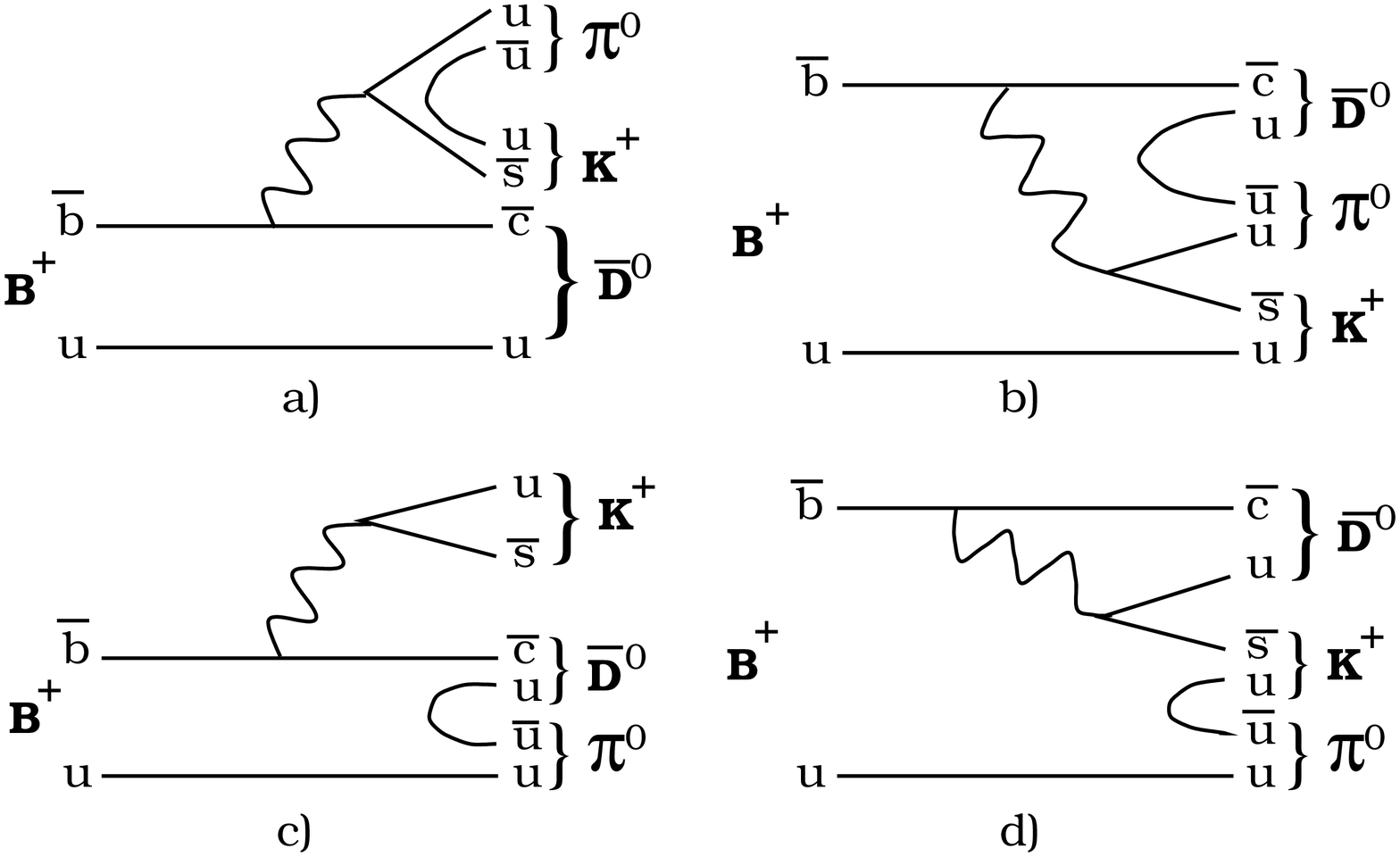,height=6.0cm,width=7.5cm}}
\end{center}
\caption{Feynmann Diagrams for the decay $\bptodzb$
 involving the CKM matrix element product $V_{cb}^*V_{us}$}
\label{fig:BDKPi1}
\end{minipage}
\end{minipage}
\hspace{1.0cm}
\begin{minipage}{.45\linewidth}
\begin{minipage}[h]{1.0\textwidth}
\begin{center}
\mbox{\epsfig{file=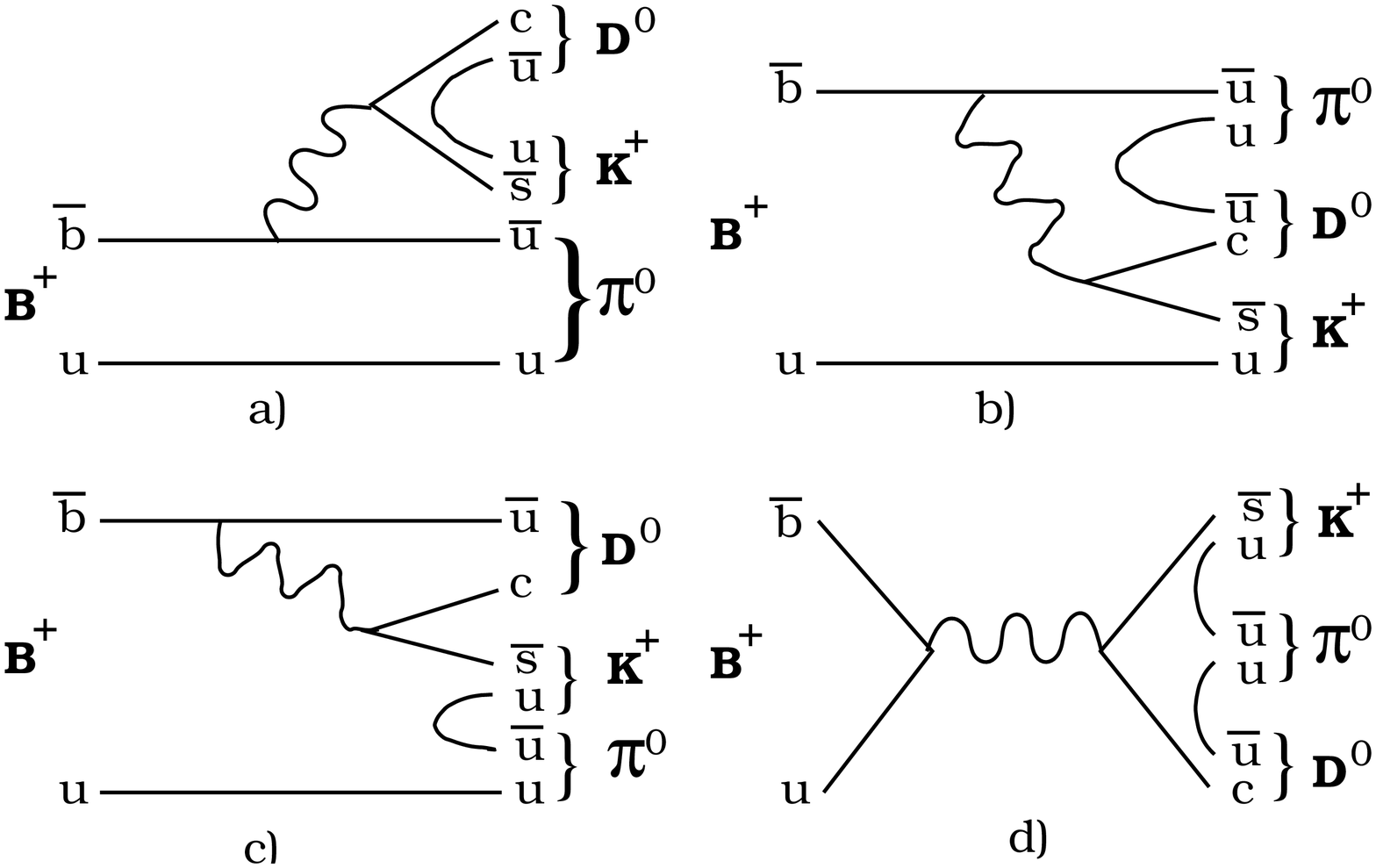,height=6.0cm,width=7.5cm}}
\end{center}
\caption{Feynmann Diagrams for the decay $\bptodz$
 involving the CKM matrix element product $V_{ub}^*V_{cs}$}
\label{fig:BDKPi2}
\end{minipage}
\end{minipage}
\vfill
\end{figure}

Figs.~\ref{fig:BDKPi1} and~\ref{fig:BDKPi2} show the diagrams leading
to the final states of interest.  As can be seen, the leading
diagrams (Fig.~\ref{fig:BDKPi1}a and~\ref{fig:BDKPi2}a) are both
color-allowed and of order $\lambda^3=\sin(\theta_c)$ in the
Wolfenstein parameterization~\cite{LW:1}, where $\theta_c$ is the
Cabibbo mixing angle. 
Due to the absence of color suppression, both interfering amplitudes
are large, avoiding the complications which arise due to the small
magnitude of the $\bbartoubar$ amplitude in the two-body decays. As a
result, observable CP-violating effects in the three-body decays are
expected to be large, and the $\bbartoubar$ amplitude is more easily
measured from the relatively large branching fraction ${\cal
B}(\bptodz)$, which is now subject to significantly less contamination
from doubly Cabibbo-suppressed $D$ meson decays than the corresponding
two-body modes.
However, should the $\bbartoubar$ amplitude be unexpectedly small, one
could still carry out the analysis described in this paper by taking
doubly Cabibbo-suppressed decays into
account~\cite{ref:soffer-ambig,ref:ADS}.

\begin{figure}[htb]
\vfill
\begin{minipage}{.45\linewidth}
\begin{minipage}[h]{1.0\textwidth}
\begin{center}
\mbox{\epsfig{file=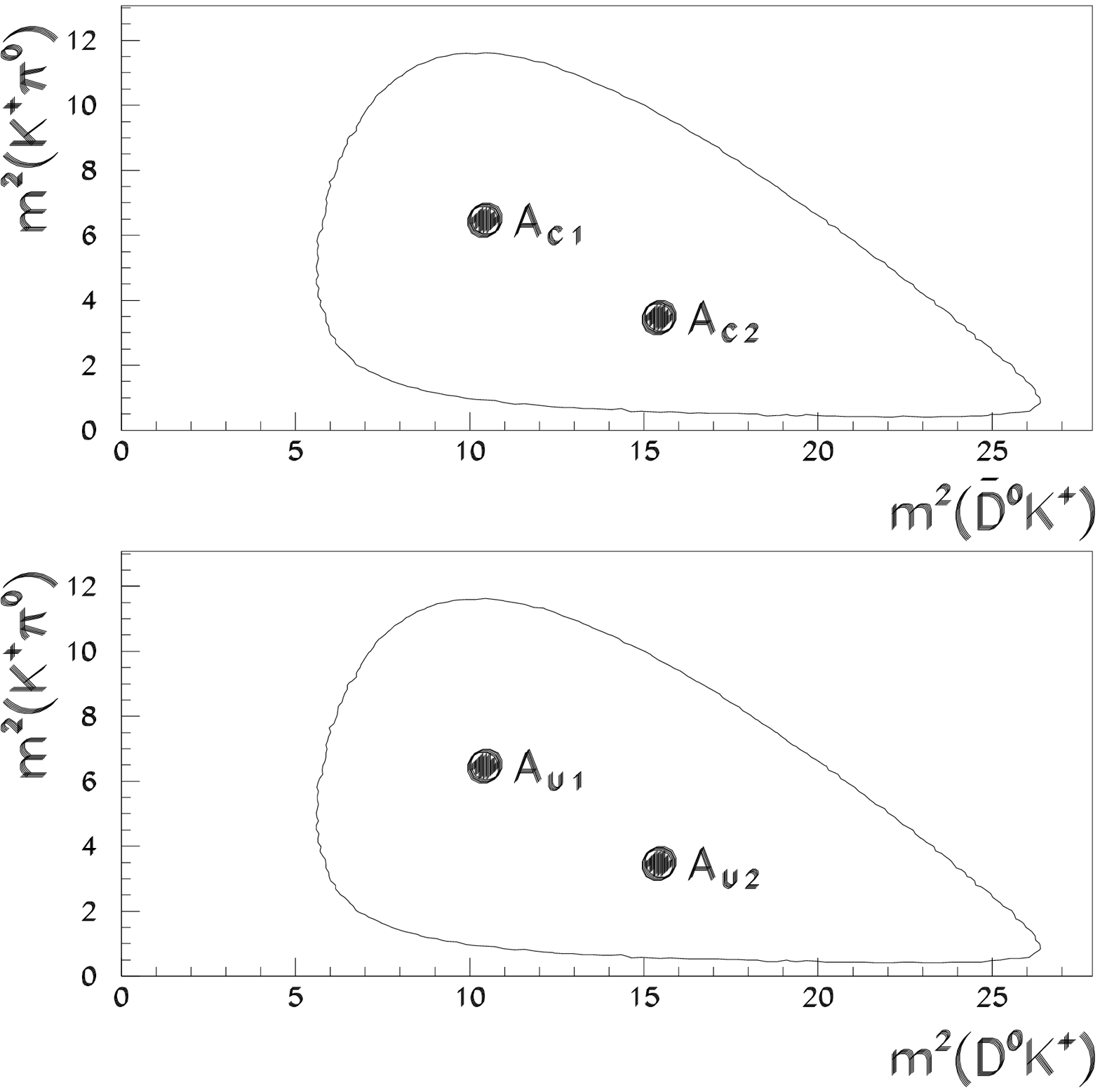,height=8.0cm,width=8.0cm}}
\end{center}
\caption{Two points on the Dalitz plot of the decays $\bptodzb$ and $\bptodz$.}
\label{fig:dalitz1}
\end{minipage}
\end{minipage}
\hspace{1.0cm}
\begin{minipage}{.45\linewidth}
\begin{minipage}[h]{1.0\textwidth}
\begin{center}
\mbox{\epsfig{file=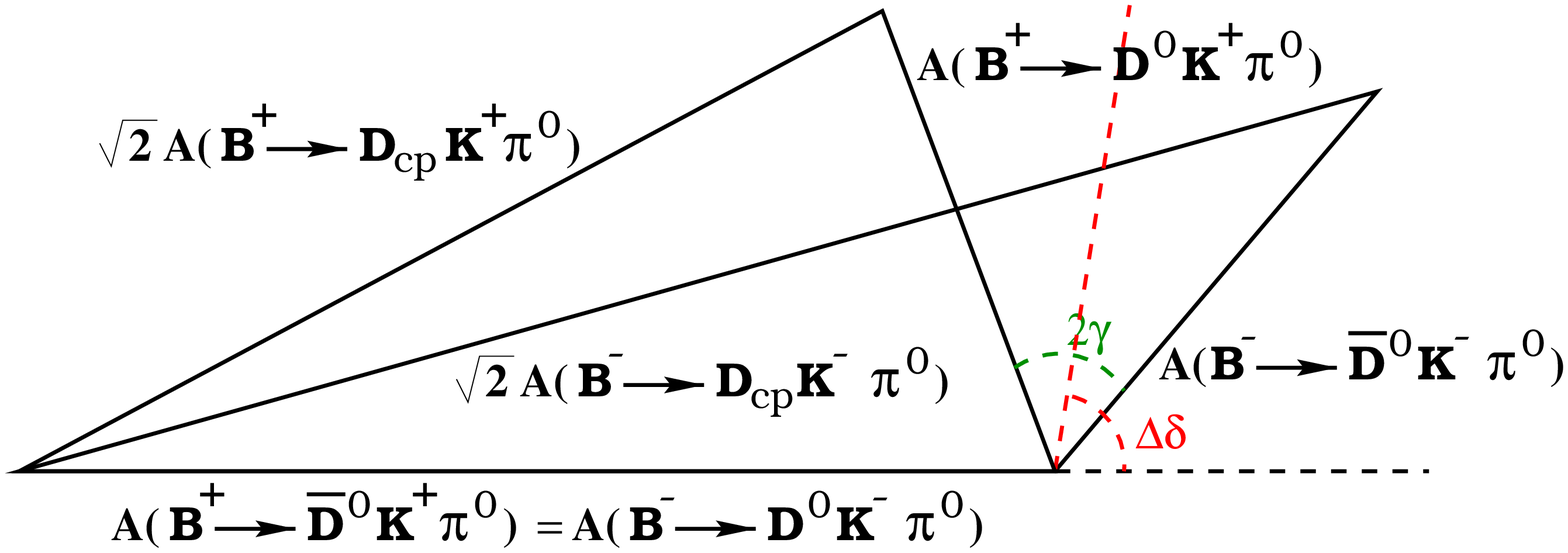,height=4.0cm,width=6.0cm}}
\end{center}
\caption{Illustration of the triangle relations in the decays $\bpmtodzb$ 
and $\bpmtodz$.}
\label{fig:triangles}
\end{minipage}
\end{minipage}
\vfill
\end{figure}

Let us examine how one could observe CP violation and measure the
angle $\gamma$ in these decays. We first consider the case of very
large statistics, and then discuss how one would proceed when the data
sample is limited.  Since we are dealing with a three-body decay, we
use the Dalitz plot of the system $DK^\pm \pi^0$ (see
Fig.~\ref{fig:dalitz1}).
Selecting a particular point $i$ in this representation, 
Eq.~(\ref{eq:D12}) implies the relations
\begin{eqnarray}
{\cal A}_i(\bptodcp) &=& {1 \over \sqrt{2}} \left(
  {\cal A}_i(\bptodz ) \pm {\cal A}_i(\bptodzb)\right) \nonumber\\[0.05truein]
{\cal A}_i(\bmtodcp) &=& {1 \over \sqrt{2}} \left(
  {\cal A}_i(\bmtodz ) \pm {\cal A}_i(\bmtodzb)\right).
\label{eq:dkpi-triangle}
\end{eqnarray}
Let us write the amplitudes corresponding to the transitions in 
Fig.~\ref{fig:BDKPi1} and~\ref{fig:BDKPi2} as
\begin{eqnarray}
{\cal A}_i(\bptodzb) = A\bc_{i}e^{i\delta\bc_{i}} \ \ \ &,& \ \ \ 
{\cal A}_i(\bptodz ) = A\bu_{i}e^{i\delta\bu_{i}} e^{i\gamma}, \nonumber\\
{\cal A}_i(\bmtodz) = A\bc_{i}e^{i\delta\bc_{i}} \ \ \ &,& \ \ \ 
{\cal A}_i(\bmtodzb ) = A\bu_{i}e^{i\delta\bu_{i}} e^{-i\gamma}, 
\label{eq:bpamp}
\end{eqnarray}
where $\gamma$ is the relative phase of the CKM matrix elements
involved in this decay, and $A\bc$ ($A\bu$) and $\delta\bc$
($\delta\bu$) are the real amplitude and CP-conserving strong
interaction phase of the transitions of Fig.~\ref{fig:BDKPi1}
(Fig.~\ref{fig:BDKPi2}). Let us note here that we have used the Wolfenstein
parameterization at order ${\cal O}$($\lambda^3$). Should one use the full 
expansion, a small weak phase of order $\lambda^4$ would be present.
Indeed the angle which is measured is
$\gamma^\prime = \arg(V_{ud} V_{ub}^\ast / V_{cs}V_{cb}^\ast)=\gamma +\xi$,
where $\xi=\arg{\left(- V_{cd} V_{cs}^\ast/ V_{ud}V_{us}^\ast\right)}$. 
The angle $\xi$ is one of the angles~\cite{ref:AKL:1}
arising from the unitarity relation
$V_{ud}V_{us}^\ast + V_{cd}V_{cs}^\ast  + V_{td}V_{ts}^\ast =0$.
The amplitudes in Eqs.~\ref{eq:bpamp} can be obtained from the
measurements of the B decay widths
\begin{eqnarray}
\Gamma_i(\bptodzb ) &=& \Gamma_i(\bmtodz ) = A\bc_{i}^2 \nonumber\\
\Gamma_i(\bptodz ) &=& \Gamma_i(\bmtodzb ) = A\bu_{i}^2.
\label{eq:width}
\end{eqnarray}
Eq.~(\ref{eq:dkpi-triangle}) implies
\begin{eqnarray}
2\Gamma_i(\bptodcp ) &=& A\bc_{i}^2 + A\bu_{i}^2 \pm 
2A\bc_{i}A\bu_{i}\cos(\Delta\delta_i+\gamma) \nonumber\\
2\Gamma_i(\bmtodcp ) &=& A\bc_{i}^2 + A\bu_{i}^2 \pm
2A\bc_{i}A\bu_{i}\cos(\Delta\delta_i-\gamma),
\label{eq:widthcp}
\end{eqnarray}
where $\Delta\delta_i \equiv \delta\bu_{i} - \delta\bc_{i}$.
Thus, by measuring the widths in Eq.~\ref{eq:width} and~\ref{eq:widthcp},
one extracts $\sin^2\gamma$ from
\begin{equation}
\sin^2\gamma = {1 \over 2} \left(1-C\overline{C} \pm 
\sqrt{(1-C^2)(1-\overline{C}^2)} \right),
\label{eq:sing_sq}
\end{equation}
where $C \equiv \cos(\Delta\delta_i+\gamma)$ and
$\overline{C}=\cos(\Delta\delta_i-\gamma)$.  Hence in the limit of
very high statistics, one would extract $\sin^2\gamma$ for each point
$i$ of the Dalitz plot and therefore obtain many measurements of the
same quantity. This would allow one to obtain a large set of redundant
measurements from which a precise and consistent value of
$\sin^2\gamma$ could be extracted. 
We note that in our treatment we disregard doubly Cabibbo-suppressed
$D^0$ decays~\cite{ref:ADS} since, due to the lack of
color-suppression, their effect is small, and can be dealt
with~\cite{ref:soffer-ambig} in any case.

In every point of the Dalitz plot, $\gamma$ is obtained with an
eight-fold ambiguity, which is a consequence of the invariance of the
$\cos(\Delta\delta_i \pm \gamma)$ terms in Eq.~(\ref{eq:widthcp})
under the three symmetry operations~\cite{ref:soffer-ambig}
\begin{equation}
\begin{array}{lllll}
\sex &:& \gamma \rightarrow \delta &, & \delta \rightarrow \gamma\\
\ssign &:&\gamma \rightarrow -\gamma &,& \delta \rightarrow -\delta  \\
\spi &:& \gamma \rightarrow \gamma + \pi &,& 
	\delta \rightarrow \delta + \pi .
\label{eq:ambig}
\end{array}
\end{equation}
However, an important benefit is gained from the multiple measurements
made in different points of the Dalitz plot. 
When results from the different points are combined, some
of the ambiguity will be resolved, in the likely case that the strong
phase $\Delta\delta_i$ varies from one region of the Dalitz to the
other. This variation can either be due to the presence of resonances or
because of a varying phase in the non-resonant contribution.
In this case, the exchange symmetry $\sex$ is numerically different
from one point to the other, which in effect breaks this symmetry and
resolves the ambiguity.

Similarly, the $\ssign$ symmetry is broken if there exists some
{\it a-priori} knowledge of the dependence of $\Delta\delta_i$ on the Dalitz
plot parameters. This knowledge is provided by the existence of broad
resonances, whose Breit-Wigner phase variation is known and may be
assumed to dominate the phase variation over the width of the
resonance.
To illustrate this, let $i$ and $j$ be two points in the Dalitz plot,
corresponding to different values of the invariant mass of the decay
products of a particular resonance. For simplicity we consider only
one resonance. One then measures $\cos(\Delta\delta_i \pm \gamma)$ at
point $i$ and $\cos(\Delta\delta_i + \alpha_{ij} \pm \gamma)$ at point
$j$, where $\alpha_{ij}$ is known from the parameters of the
resonance. 
It is important to note that the sign of $\alpha_{ij}$ is also known,
hence it does not change under $\ssign$. Therefore, should one choose
the $\ssign$-related solution $\cos(-\Delta\delta_i \mp \gamma)$ at
point $i$, one would get 
$\cos(-\Delta\delta_i + \alpha_{ij} \mp \gamma)$ 
at point $j$. Since this is different from 
$\cos(\Delta\delta_i + \alpha_{ij} \pm \gamma)$,
the $\ssign$ ambiguity is resolved. 
This is illustrated graphically in Eq.~\ref{eq:ssign}:
\begin{equation}
\begin{array}{ccc}
\cos(\Delta\delta_i \pm \gamma) & \stackrel{\ssign}{\longleftrightarrow}& 
\cos(-\Delta\delta_i \mp \gamma) \\
BW \downarrow & & \downarrow BW \\
\cos(\Delta\delta_i + \alpha_{ij} \pm \gamma) 
& \;\stackrel{\ssign}{\not\!\!\!\longleftrightarrow} & 
\cos(-\Delta\delta_i + \alpha_{ij} \mp \gamma)
\end{array}
\label{eq:ssign}
\end{equation}

Thus, broad resonances reduce the initial eight-fold ambiguity to the
two-fold ambiguity of the $\spi$ symmetry, which is not broken.
Fortunately, $\spi$ leads to the well-separated solutions $\gamma$ and
$\gamma + \pi$, the correct one of which is easily identified when
this measurement is combined with other measurements of the unitarity
triangle.

\section{The Finite Statistics Case}
Since experimental data sets will be finite, extracting $\gamma$ will
require making use of a limited set of parameters to describe the
variation of amplitudes and strong phases over the Dalitz plot. The
consistency of this approach can be verified by comparing the results
obtained from fits of the data in a few different regions of the
Dalitz plot, and the systematic error due to the choice of the
parameterization of the data may be obtained by using different
parameterizations.

A fairly general parameterization assumes the existence of $N_R$
Breit-Wigner resonances, as well as a non-resonant contribution:
\begin{eqnarray}
{\cal A}\,_\xi(\bptodzb) &=& \left(A\bc_{0}\, e^{i\delta\bc_{0}} + 
	\sum_{j=1}^{N_R} A\bc_{j}B_{s_j}(\xi) \, e^{i\delta\bc_{j}}\right) 
	e^{i\delta\bc(\xi)}
	\nonumber\\
{\cal A}\,_\xi(\bptodz ) &=& \left(A\bu_{0}\, e^{i\delta\bu_{0}} + 
	\sum_{j=1}^{N_R} A\bu_{j}B_{s_j}(\xi) \, e^{i\delta\bu_{j}} \right)
	e^{i\delta\bu(\xi)} e^{i\gamma},
\label{eq:amp-param}
\end{eqnarray}
where $\xi$ represents the Dalitz plot variables, 
\begin{equation}
B_{s_j}(\xi) \equiv b_{s_j}(\xi)\, e^{i\delta_j(\xi)}
\label{eq:BW}
\end{equation}
is the Breit-Wigner amplitude for a particle of spin $s_j$, normalized
such that $\int (b_{s_j}(\xi))^2 d\xi = 1$,
$A\bu_0$ and $\delta\bu_0$ ($A\bc_0$ and $\delta\bc_0$) are the
magnitude and CP-conserving phase of the non-resonant $\bbartoubar$
($\bbartocbar$) amplitude, 
and $A\bu_j$ and $\delta\bu_j$ ($A\bc_j$ and $\delta\bc_j$) are the
magnitudes and CP-conserving phase of the $\bbartoubar$
($\bbartocbar$) amplitude associated with resonance $j$~\cite{ref:D-Dalitz}.
The functions $\delta\bc(\xi)$ and $\delta\bu(\xi)$ may be assumed to
vary slowly over the Dalitz plot, allowing their description in terms
of a small number of parameters. 
Eq.~(\ref{eq:D12}) again implies 
\begin{equation}
{\cal A}\,_\xi(\bptodcp) = {1 \over \sqrt{2}} \biggl(
	{\cal A}\,_\xi(\bptodz ) \pm {\cal A}\,_\xi(\bptodzb) \biggr).
\label{eq:amp-D12}
\end{equation}
The decay amplitudes of $B^-$ mesons are identical to those of
Eqs.~(\ref{eq:amp-param}) and~(\ref{eq:amp-D12}), with $\gamma$ replaced
by $-\gamma$.

The decay amplitudes of Eqs.~(\ref{eq:amp-param}) and~(\ref{eq:amp-D12}) can
be used to conduct the full data analysis. This is done by
constructing the probability density function (PDF)
\begin{equation}
P(\xi) = |{\cal A}\,_{\xi}(f)|^2, 
\label{eq:PDF}
\end{equation}
where the amplitude $A\,_{\xi}(f)$ is given by one of the expressions
of Eq.~(\ref{eq:amp-param}), Eq.~(\ref{eq:amp-D12}), or their
CP-conjugates, depending on the final state $f$. Given a sample of
$N_e$ signal events, $\gamma$ and the other unknown parameters of
Eq.~(\ref{eq:amp-param}) are determined by minimizing the negative log
likelihood function
\begin{equation}
\chi^2 \equiv -2 \sum_{i=1}^{N_e} \log P(\xi_i),
\label{eq:chi2}
\end{equation}
where $\xi_i$ are the Dalitz plot variables of event $i$.

\section{Resonances and Ambiguities}
\label{sec:res}
It is worthwhile to consider the resonances which may contribute 
to the $\dzkpmpiz$ final state.
Obvious candidates are broad $D^{**}$ and $D_s^{**}$
states. However, only the ones which can decay as
$D^{**0} \rightarrow D^0 \pi^0$ 
or 
$D_s^{**+} \rightarrow D^0 K^+$ 
are relevant for the final state of interest. This exclude the $1^+$
states, which would decay to $D^* \pi$ or $D^* K$. Furthermore, since
the $D_s^{**+}$ is essentially produced through a $W^+$, the $2^+$
state is forbidden as well. 
Thus, one does not expect a large contribution from these states.
A promising candidate might be the broad $D^{*0}_0$ recently
observed by the Belle collaboration~\cite{ref:belle-D2star}.
We note that including such resonances
in the analysis does not raise particular difficulties and 
would further enhance the sensitivity of the $\gamma$ measurement.
Similar arguments can be made for higher excited $K$ states.

\begin{figure}[htb!]
\begin{center}
\epsfig{file=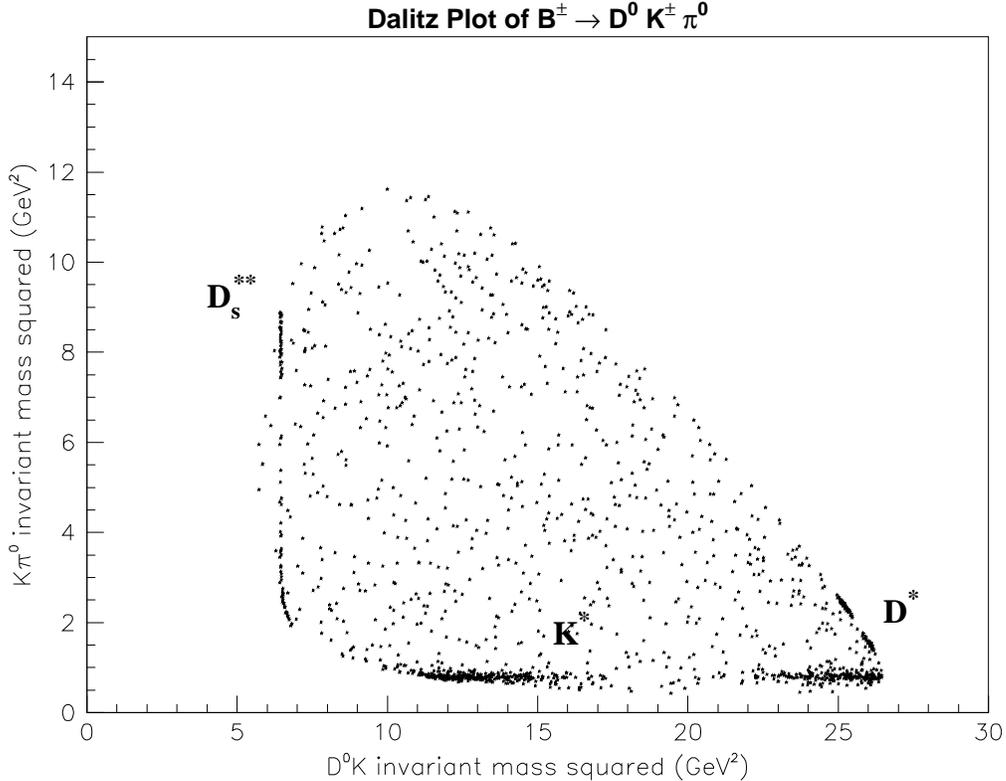,height=120mm}
\caption{Dalitz plots obtained from a simulation of $B^+$ and $B^-$ decays 
into all final state,
$D^0 K^{\pm} \pi^0$,
$\overline{D^0} K^{\pm} \pi^0$, and
$D_{1,2} K^{\pm} \pi^0$. 
with the parameters of Table~\ref{tab:reference}. Along with non-resonant
contributions, the resonances $\kstar^0$, $D^{* 0}$, and $D_s^{**}$
are shown.}
\label{fig:dalitz}
\end{center}
\end{figure}

One also expects narrow resonances, such as the $D^{*}(2007)^0$ and a
narrow $D_s^{**+}$ state, to be produced. However, as seen in the
Dalitz plot of Fig.~\ref{fig:dalitz}, these resonances do not overlap,
and hence do not interfere. In addition, interference between a very
narrow resonance and either a broad resonance or a non-resonant term is
suppressed in proportion to the square root of the narrow resonance
width.  Therefore, narrow resonances contribute significantly to the
CP violation measurement only if both the $\bbartocbar$ and
$\bbartoubar$ amplitudes proceed through the same resonance. This
scenario is favorable, but is not necessary for the success of
our method, and will therefore not be focused on in the rest of this
study.

In what follows, we discuss important properties of the method by
considering the illustrative case, in which the $\bbartoubar$ decay
proceeds only via a non-resonant amplitude, and the $\bbartocbar$
decay has a non-resonant contribution and a single resonant
amplitude. For concreteness, the resonance is taken to be the
$\kstar{^\pm}(892)$. We take the $\xi$-dependent non-resonant
phases to be $\delta\bc(\xi) = \delta\bu(\xi) = 0$.
Under these circumstances, the PDF of Eq.~(\ref{eq:PDF}) depends on
four cosine terms that are measured in the experiment:
\begin{eqnarray}
c_{00}^\pm &\equiv& \cos(\delta\bu_0 \pm \gamma) \nonumber\\
c_{\kstar 0}^\pm &\equiv& \cos(\delta\bu_0 - \delta_{\kstar}  
	- \delta_\kstar(\xi) \pm \gamma),
\label{eq:4-cosines}
\end{eqnarray}
where $\delta_\kstar(\xi)$ is the $\xi$-dependent $\kstar$ Breit-Wigner
phase of Eq.~(\ref{eq:BW}).
The cosines $c_{00}^\pm$ ($c_{\kstar 0}^\pm$) arise from interference
between the non-resonant (resonant) $\bbartocbar$ amplitude and the
non-resonant $\bbartoubar$ amplitude.

The phases $\delta\bu_0$, $\delta_{\kstar}$, and $\gamma$ are all {\it
a-priori} unknown. However, it is important to note that $\delta_{\kstar}$ is
fully determined from the interference between the resonant and
non-resonant contributions to the relatively high statistics decay
mode $B^+\rightarrow \overline D^0 K^+ \pi^0$ as a function of the
Dalitz plot variables. 
Therefore, $\delta_{\kstar}$ is obtained with
no ambiguities, and with an error much smaller than those of
$\delta\bu_0$ or 
$\gamma$\footnote{
	Even when one of the $\bbartocbar$ amplitudes
	is small enough that the determination of $\delta_{\kstar}$ becomes
	difficult, one effectively has
	$\delta\bu_0 \rightarrow \delta\bu_0 - \delta_{\kstar}$,
	$\delta_{\kstar} \rightarrow 0$,
	and the determination of $\delta_{\kstar}$ is again not a problem.}.
Consequently, the only relevant symmetry operations are
\begin{equation}
\begin{array}{lllll}
\sex &:& \gamma \rightarrow \delta\bu_0 &, & \delta\bu_0 \rightarrow \gamma \\
\ssign &:&\gamma \rightarrow -\gamma &,& 
	\delta\bu_0 \rightarrow -\delta\bu_0 \\
\spi &:& \gamma \rightarrow \gamma + \pi &,& 
	\delta\bu_0 \rightarrow \delta\bu_0 + \pi \\
\sexKst&:& \gamma \rightarrow \delta\bu_0 - \delta_{\kstar} &,&
	\delta\bu_0 \rightarrow \gamma + \delta_{\kstar} \\
\sexKstm&:& \gamma \rightarrow -\delta\bu_0 + \delta_{\kstar} &,&
	\delta\bu_0 \rightarrow -\gamma + \delta_{\kstar} .
\end{array}
\label{eq:ambig-rnr}
\end{equation}
As discussed above, only $\spi$ is a symmetry of all four cosines of
Eq.~(\ref{eq:4-cosines}), and is therefore fully unresolved. 
The transformation properties of the cosines under any combination of
the remaining four operations that can lead to an ambiguity are shown
in Table~\ref{tab:cosines}.
\begin{table}[!htb]
\begin{center}
\caption{Invariance of each of the cosines of Eq.~(\ref{eq:4-cosines})
	under combinations of the symmetry operations of 
	Eq.~(\ref{eq:ambig-rnr}), excluding $\spi$. 
	Full invariance (approximate invariance) is indicated by a 
	\inv (\pinv).}
\label{tab:cosines}
\begin{tabular}{lcccc}
Operation & $c_{\kstar 0}^+$ & $c_{\kstar 0}^-$ & $c_{00}^+$ & $c_{00}^-$ \\
\hline
\multicolumn{5}{c}{Non-resonant regime} \\
$\sex$               & \inv   & \noinv & \inv   & \inv   \\
$\ssign$             & \noinv & \noinv & \inv   & \inv   \\
$\sex\ssign$         & \noinv & \inv   & \inv   & \inv   \\
\hline
\multicolumn{5}{c}{Resonant regime} \\
$\sexKst$            & \inv   & \pinv  & \inv   & \noinv \\
$\sexKstm$           & \pinv  & \inv   & \noinv & \inv   \\
$\sexKst\sexKstm$    & \pinv   & \pinv & \noinv & \noinv \\
\end{tabular}
\end{center}
\end{table}

While none of the operations leaves all four cosines invariant, it is
important to note cases where $c_{\kstar 0}^\pm$ are {\it
approximately} invariant under $\sexKst$, $\sexKstm$, or their
product. We define approximate invariance under the operation $S$ to be 
\begin{equation}
S_{\rm app} \, c_{\kstar 0}^\pm( \delta_\kstar(\xi))
  = c_{\kstar 0}^\pm(-\delta_\kstar(\xi)).
\label{eq:app}
\end{equation}
Approximate invariance arises due to the fact that far from the peak
of the $\kstar$ resonance, $\delta_\kstar(\xi)$ changes slowly as a
function of the $K\pi$ invariant mass, and takes values around
$0$ and $\pi$. Therefore, for events in the tails of the Breit-Wigner,
$\delta_\kstar(\xi)$ is almost invariant under any $S_{\rm app}$
satisfying Eq.~(\ref{eq:app}).
One can see that approximate invariance of one of the cosines
$c_{\kstar 0}^\pm$ implies minimal change in the $\chi^2$ of
Eq.~(\ref{eq:chi2}), which may result in a resolved yet clearly
observable ambiguity.  Since both $c_{\kstar 0}^\pm$ terms are only
approximately invariant under the product $\sexKst\sexKstm$, this
ambiguity is more strongly resolved than either $\sexKst$ or
$\sexKstm$.

Observing that no single operation in the Table~\ref{tab:cosines} is a
good symmetry of all cosines, one identifies two different regimes: 
In the non-resonant regime, interference with the non-resonant
$\bbartocbar$ is dominant, and only $\sex$ and $\ssign$ may lead to
ambiguities.
In the resonant regime, the $\kstar$ amplitude strongly dominates the
$\bbartocbar$ decay, and $\sexKst$ and $\sexKstm$ become the
important ambiguities. 
In the transition between these regimes, the operations of
Table~\ref{tab:cosines} do not lead to clear ambiguities, as we have
verified by simulation (See sec.~\ref{sec:simulation}). Thus, while
naively one may expect a $2^5$-fold ambiguity, in practice the
observable ambiguity is no larger than eight-fold, with only the
two-fold $\spi$ being fully unresolved, in the likely case of
non-negligible resonant contribution. This is demonstrated in
Fig.~\ref{fig:scan-rnr}.
Furthermore, although one may write down more products of the operations 
$\sex$, $\ssign$, $\sexKst$, and $\sexKstm$, only the products listed
in Table~\ref{tab:cosines} result in full or partial invariance of 
both cosines which dominate the same regime. The additional products
do not result in any noticeable ambiguities.

\section{Measurement Sensitivity and Simulation Studies}
\label{sec:simulation}
To study the feasibility of the analysis using Eq.~(\ref{eq:chi2})
and verify the predictions of Sec.~\ref{sec:res}, we
conducted a simulation of the decays
$B^{\pm} \rightarrow D^0 K^{\pm} \pi^0$,
$B^{\pm} \rightarrow \overline{D^0} K^{\pm} \pi^0$, and
$B^{\pm} \rightarrow D_{1,2} K^{\pm} \pi^0$. 
Events were generated according to the PDF of Eq.~(\ref{eq:PDF}), with
the base parameter values given in Table~\ref{tab:reference}.
In this table and throughout the rest of the paper, we use a tilde to
denote the ``true'' parameter values used to generate events, while
the corresponding plain symbols represent the ``trial'' parameters
used to calculate the experimental $\chi^2$.

The only non-vanishing amplitudes in the simulation were the
non-resonant amplitudes in the $\bbartocbar$ and $\bbartoubar$ decays,
and the $\kstar$ resonant $\bbartocbar$ amplitude.
For simplicity, additional resonances were not included in this
demonstration. However, broad resonances that are observed in the
data should be included in the actual data analysis. 
%

\begin{table}[ht!]
\begin{center}
\caption{Parameters used to generate events in the simulation.
	The value of $\tilde A\bc_{\kstar}$ is chosen so as to 
	roughly agree with the measurement of the corresponding branching 
	fraction~\protect\cite{ref:cleo-btodkst}, taking into account
	the $\kstar{^+}\rightarrow K^+\pi^0$ branching fraction.}
\label{tab:reference}
\begin{tabular}{cc|cc}
  Parameter              &Value   &Parameter            &Value \\
\hline
$\tilde\gamma$            &1.20 & $\tilde A\bu_0 / \tilde A\bc_0$      &0.4\\
$\tilde{\delta}\bc(\xi) = \tilde{\delta}\bu(\xi)$ &0 & 
	$\tilde A\bc_{\kstar} / \tilde A\bc_0 $ &1.0\\
$\tilde\delta_{\kstar}$ &1.8 & $\tilde A\bc_{\kstar}$    
	&$\sim \sqrt{2 \times 10^{-4} \Gamma_B}$\\
$\tilde\delta\bu_0$ & 0.4 & \\
\end{tabular}
\end{center}
\end{table}

The simulations were conducted with a benchmark integrated luminosity
of \lumi, which each of the asymmetric B-factories plan to collect
by about 2005.
The final state reconstruction efficiencies were calculated based on
the capabilities of current $\Upsilon(4{\rm S})$ detectors. We assumed
an efficiency of 70\% for reconstructing the $K^\pm$, including track
quality and particle identification requirements, and 60\% for
reconstructing the $\pi^0$. The product of reconstruction efficiencies
and branching fractions of the $D^0$, summed over the final states
$K^-\pi^+$, $K^-\pi^+\pi^0$, and $K^-\pi^+\pi^-\pi^+$, is taken to
yield an effective efficiency of 6\%. Using the CP-eigenstate final
states $K^+K^-$, $\pi^+\pi^-$, $K_S\pi^0$, and $K_S\rho^0$, the
effective efficiency for the sum of the $D_1$ and $D_2$ final states
is 0.8\%. All efficiencies are further reduced by a factor of 1.7, in
order to approximate the effect of background.
The numbers of signal events obtained in each of the final states with
the above efficiencies and the parameters of Table~\ref{tab:reference}
are listed in Table~\ref{tab:nsignal}.  

\begin{table}
\begin{center}
\caption{The numbers of events obtained by averaging 100 simulations using
the parameters of Table~\protect\ref{tab:reference} and the reconstruction
efficiencies listed in the text.}
\label{tab:nsignal}
\begin{tabular}{lc}
Mode & Signal events per 400\fb \\
\hline
$B^+ \rightarrow \overline{D^0} K^+\pi^0 = 
	B^- \rightarrow {D^0} K^-\pi^0$ & 2610 \\
$B^+ \rightarrow           D^0  K^+\pi^0 = 
	B^- \rightarrow \overline{D^0}  K^-\pi^0$ &  205 \\
$B^+ \rightarrow        D_{1,2} K^+\pi^0$ &  186 \\
$B^- \rightarrow        D_{1,2} K^-\pi^0$ &  234 \\
\end{tabular}
\end{center}
\end{table}

\begin{figure}[htb!]
\begin{center}
\begin{tabular}{c}
\epsfig{file=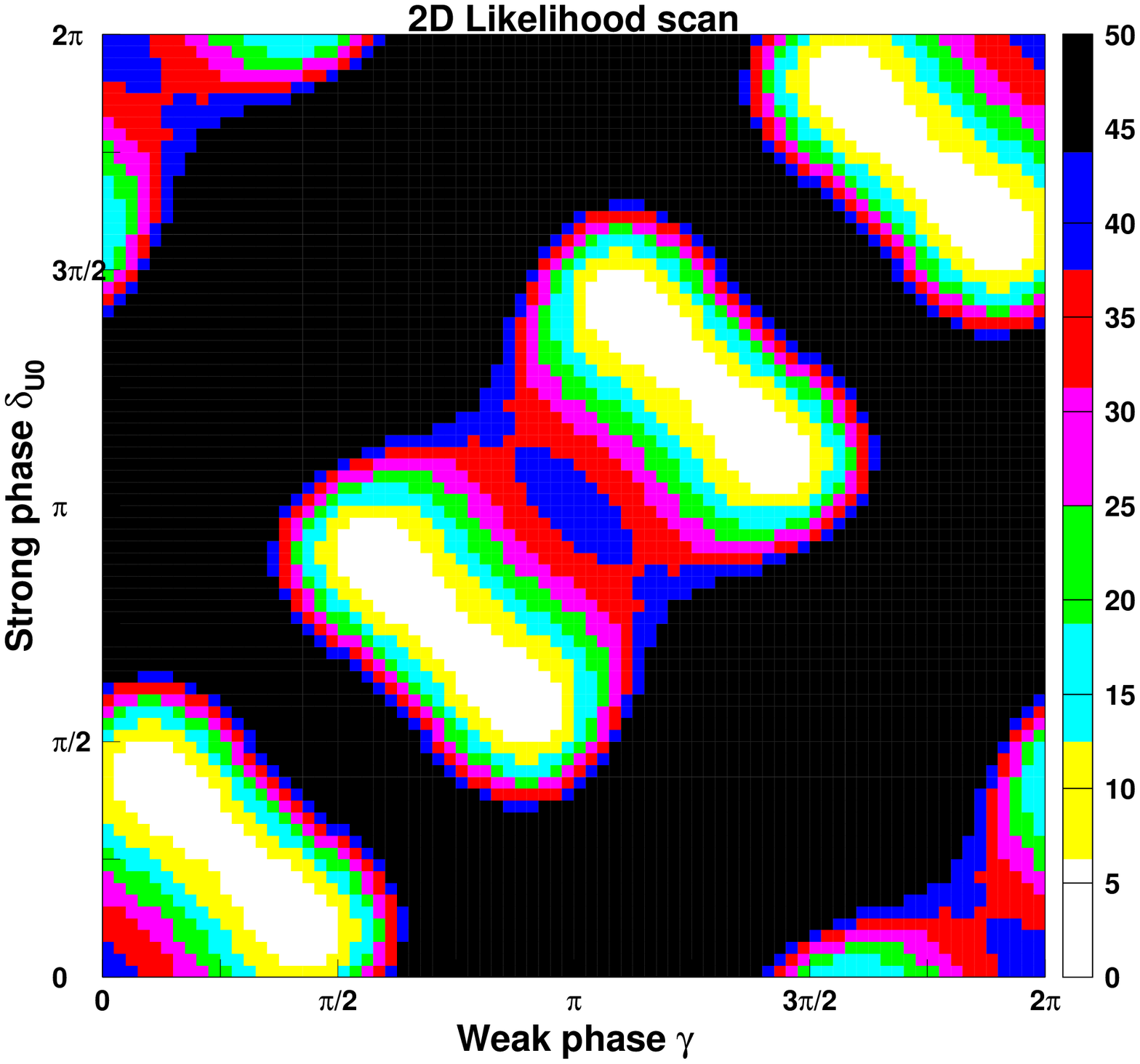,height=90mm} \\
\epsfig{file=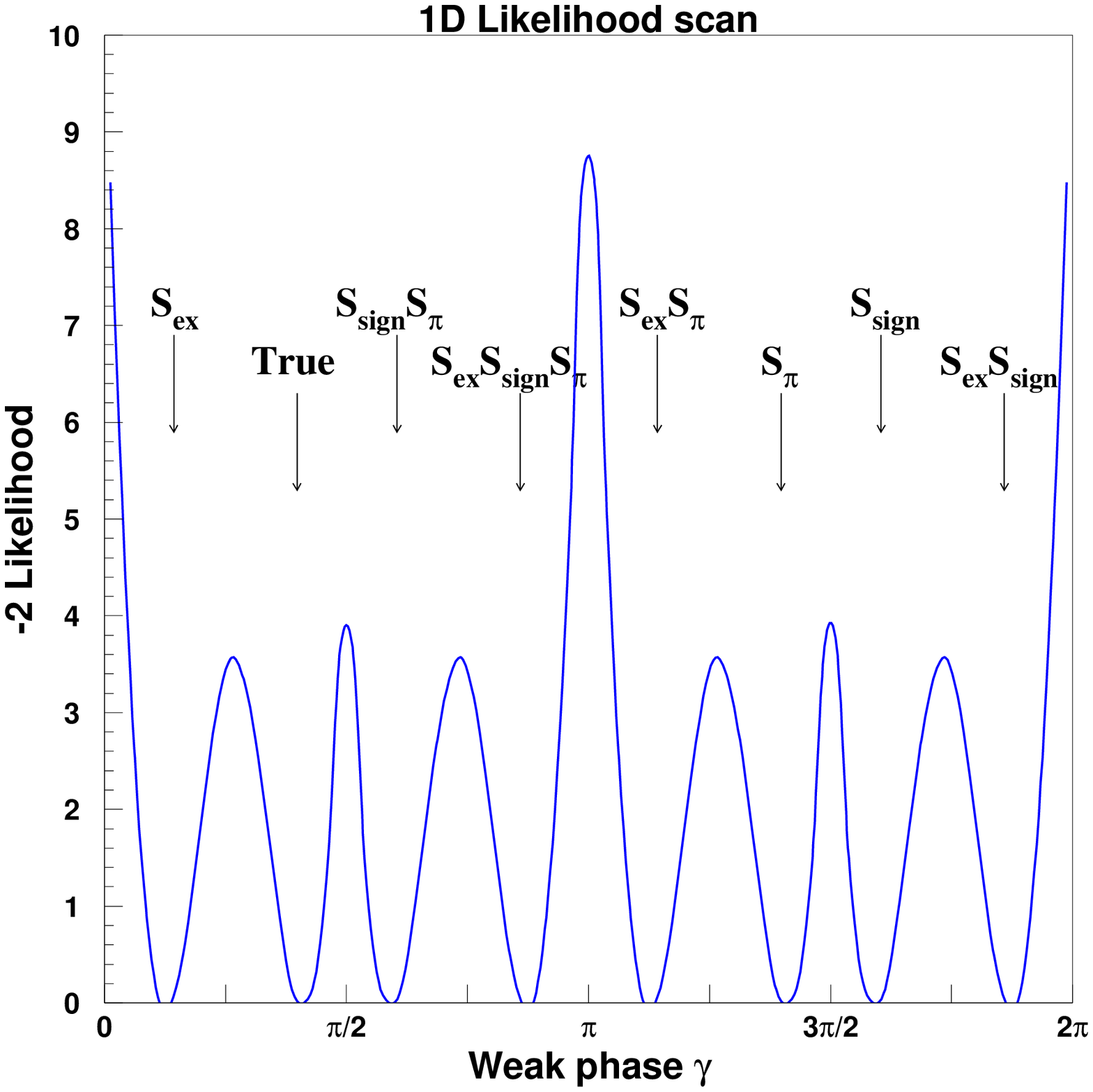,height=90mm} \\
\end{tabular}
\caption{{\bf Top:} $\chi^2$ as a function of $\gamma$ and $\delta\bu_0$, with
	the parameters of Table~\ref{tab:reference} and 
	no resonant contribution ($\tilde A\bc_{\kstar} = 0$).
	{\bf Bottom:} Minimum projection
	of $\chi^2$ onto $\gamma$.}
\label{fig:scan-nr}
\end{center}
\end{figure}

\begin{figure}[htb!]
\begin{center}
\begin{tabular}{c}
\epsfig{file=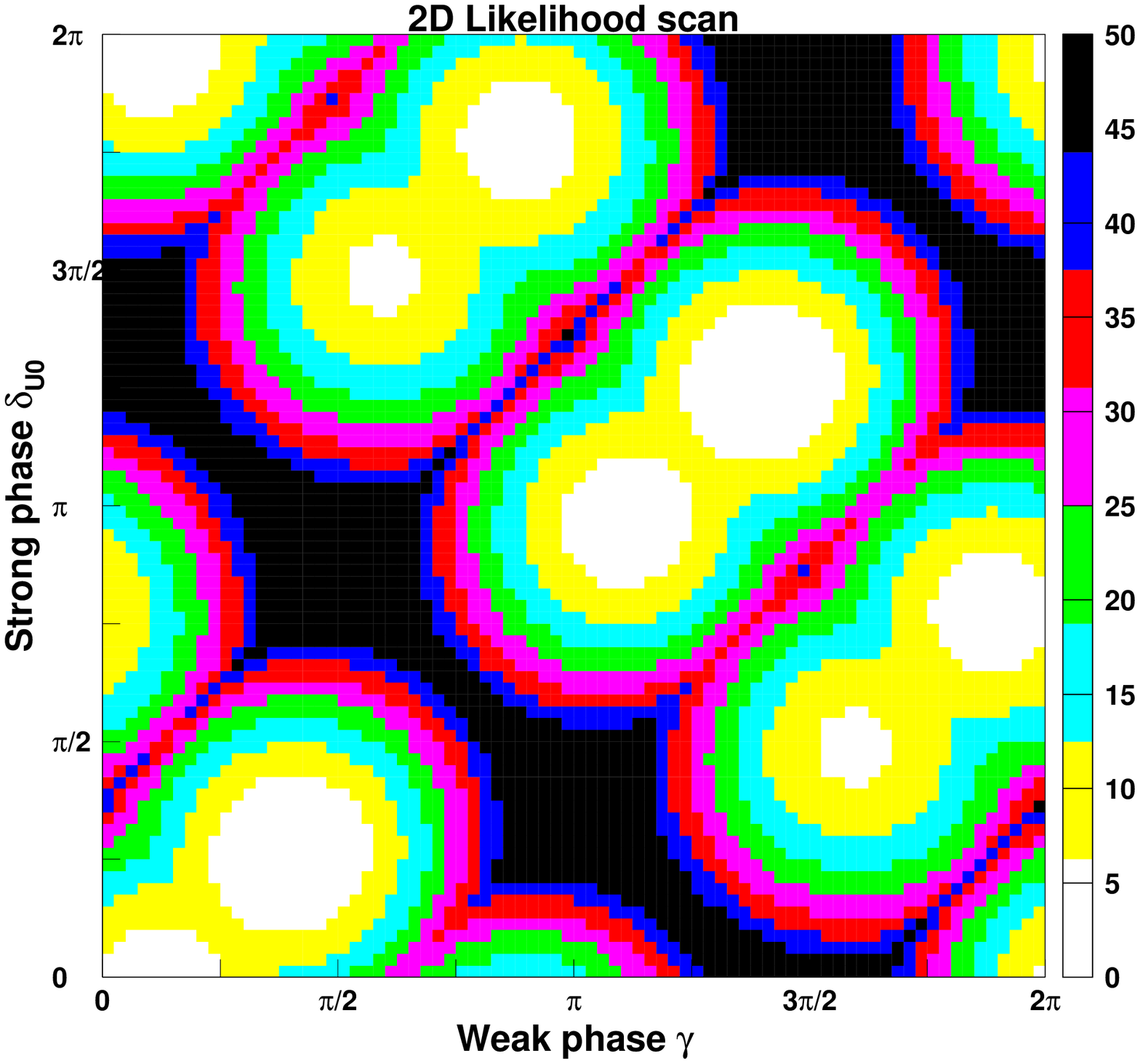,height=90mm} \\
\epsfig{file=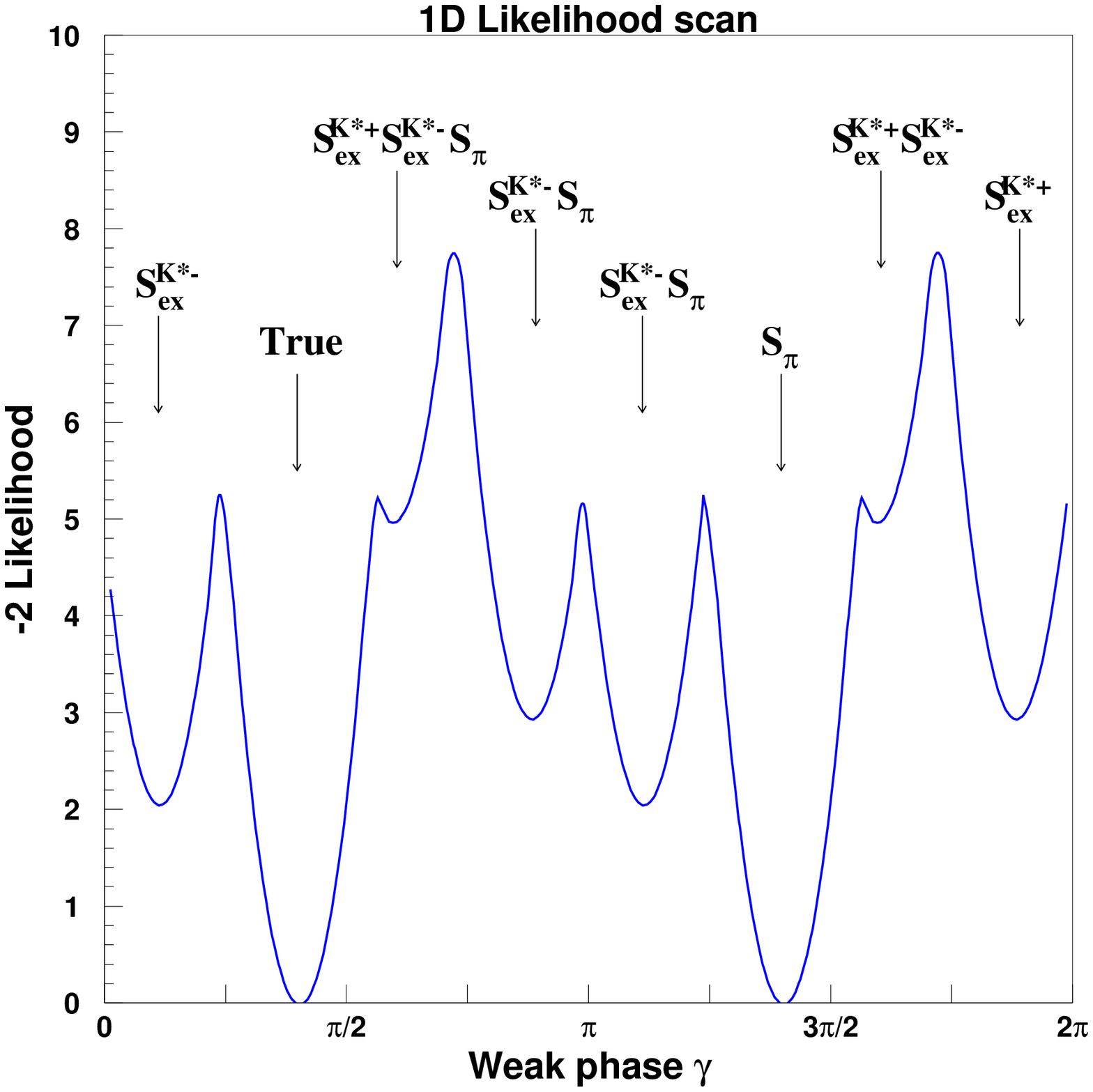,height=90mm} \\
\end{tabular}
\caption{{\bf Top:} $\chi^2$ as a function of $\gamma$ and $\delta\bu_0$, with
	no non-resonant $\bbartocbar$ contribution 
	($\tilde A\bc_0 = 0$). The value $\delta_{\kstar} = 1.2$ is used
	to ensure that ambiguities do not overlap. All other
	parameters are those of Table~\ref{tab:reference}.
	{\bf Bottom:} Minimum projection
	of $\chi^2$ onto $\gamma$.}
\label{fig:scan-r}
\end{center}
\end{figure}

\begin{figure}[htb!]
\begin{center}
\begin{tabular}{c}
\epsfig{file=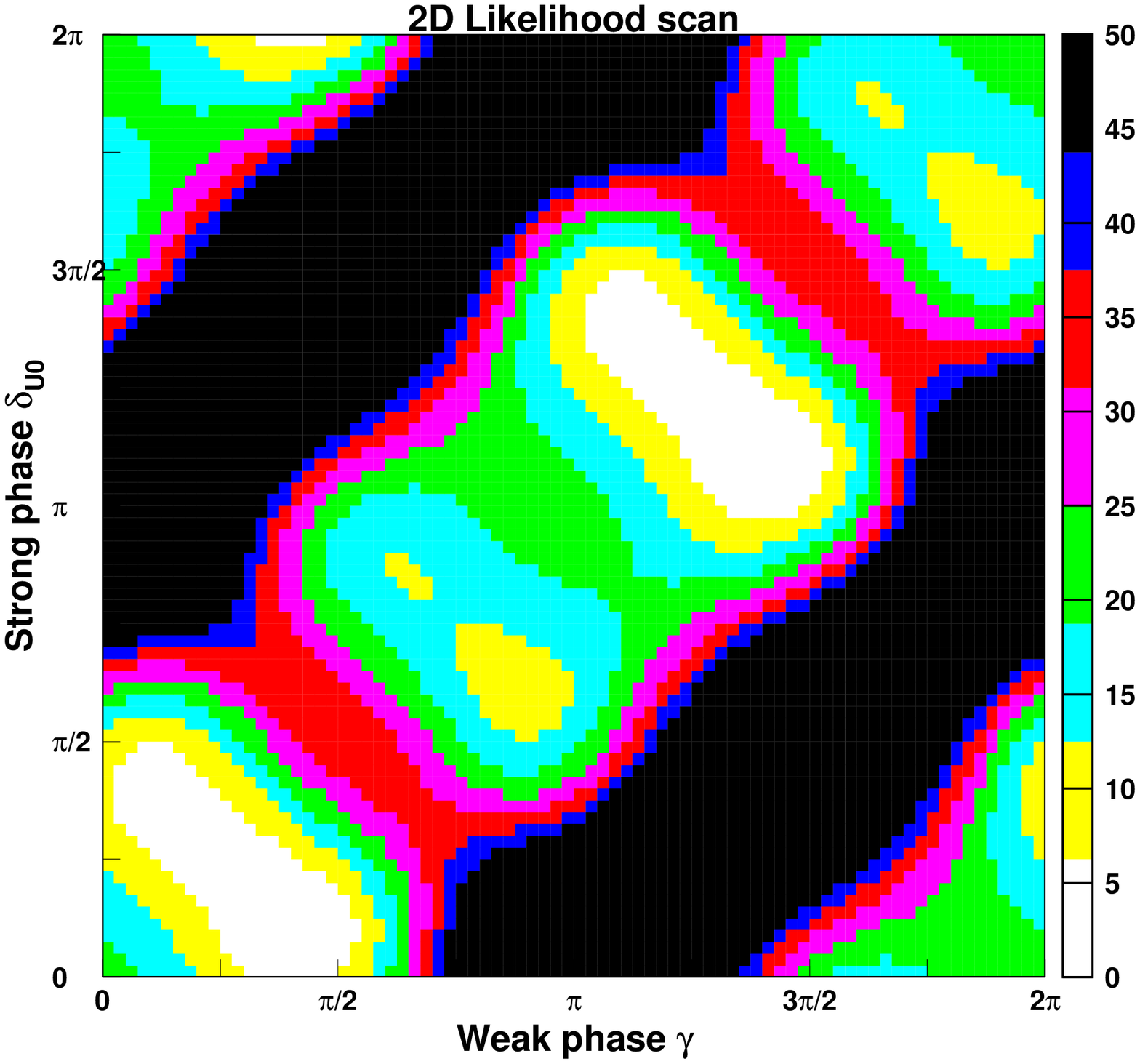,height=90mm} \\
\epsfig{file=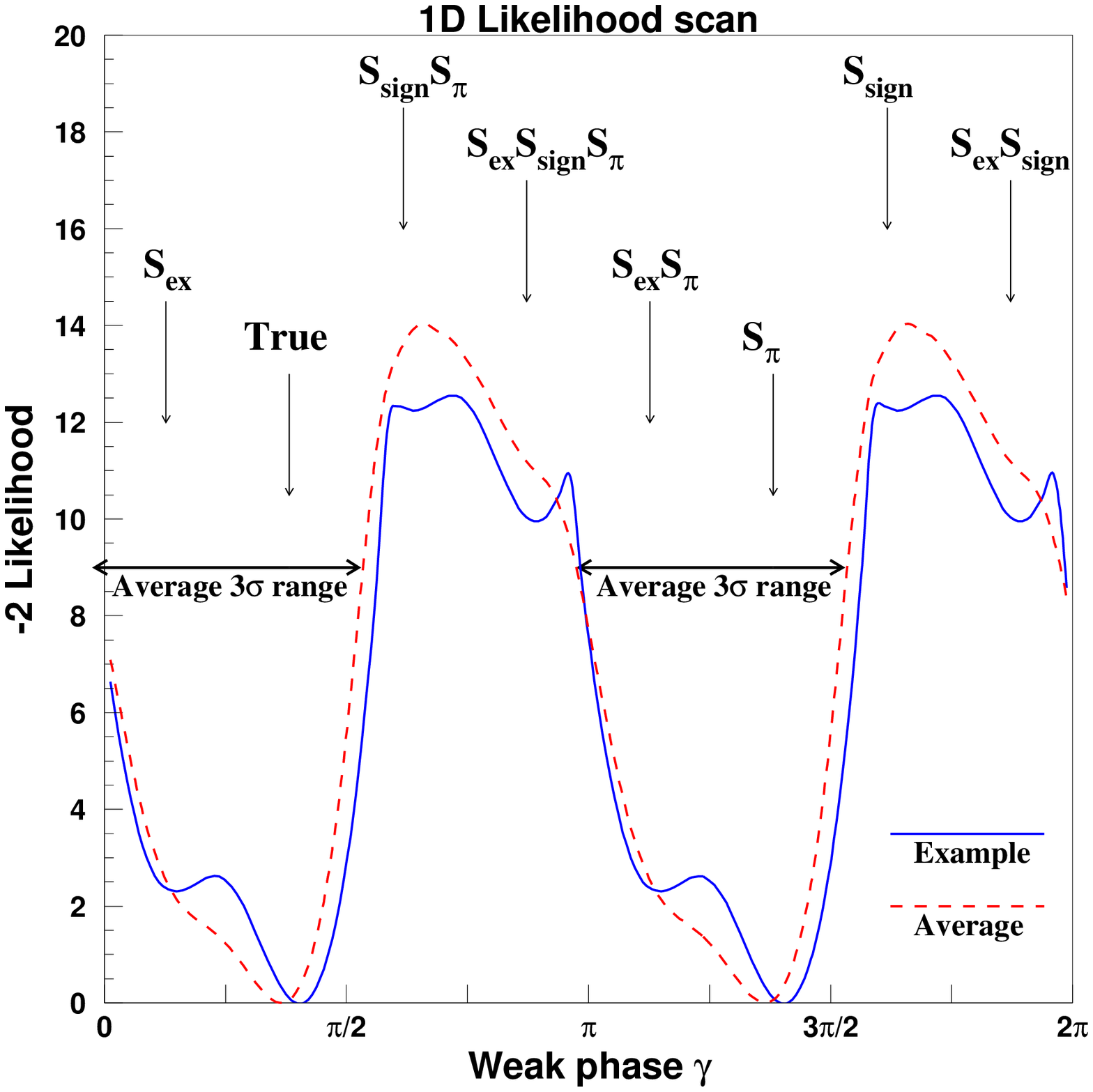,height=90mm} \\
\end{tabular}
\caption{{\bf Top:} $\chi^2$ as a function of $\gamma$ and $\delta\bu_0$, with
	the parameters of Table~\ref{tab:reference}.
	{\bf Bottom:} The solid line shows the minimum projection
	of $\chi^2$ onto $\gamma$ for the example experiment. The dashed line
	represents the average of 50 simulated experiments. The three standard
	deviation allowed range of $\gamma$ obtained from the average is 
	indicated by arrows.}
\label{fig:scan-rnr}
\end{center}
\end{figure}

In Figs.~\ref{fig:scan-nr} through~\ref{fig:scan-rnr}, we show the
dependence of $\chi^2$ on the values of $\gamma$ and
$\delta\bu_0$. The smallest value of $\chi^2$ is shown as zero.
At each point in these figures, $\chi^2$ is calculated with the
generated values of the amplitude ratios $A\bu_0 / A\bc_0 = \tilde
A\bu_0 / \tilde A\bc_0$ and $A\bc_{\kstar}/ A\bc_0 = \tilde
A\bc_{\kstar}/ \tilde A\bc_0$.  
We note that when these amplitude ratios are determined by a fit simultaneously
with the phases, the correlations between the amplitudes and the
phases are generally found to be less than 20\%.
Therefore, the results obtained with the amplitudes fixed to their
true values are sufficiently realistic for the purpose of this
demonstration.

For each of these figures, we also show the one-dimensional minimum projection
$\chi^2(\gamma)= \min\{ \chi^2(\gamma, \delta\bu_0)\}$,
showing the smallest value of $\chi^2$ for each value of $\gamma$.

Fig.~\ref{fig:scan-nr} is a simulation obtained with the parameters of
Table~\ref{tab:reference}, but with $A\bc_{\kstar} = 0$. With no
resonant contribution, the eight-fold ambiguity of the perfect
non-resonant regime is clearly visible. This would be the typical case
for two-body final states.

Fig.~\ref{fig:scan-r} is obtained with the parameters of
Table~\ref{tab:reference}, but with $A\bc_0 = 0$. With no non-resonant
$\bbartocbar$ contribution, the eight-fold ambiguity of the perfect
resonant regime is seen. The ambiguities corresponding to approximate
invariance are clearly resolved, with the doubly-approximate
$\sexKst\sexKstm$ ambiguity resolved more strongly.

Fig.~\ref{fig:scan-rnr} is obtained with the parameters of
Table~\ref{tab:reference} and shows how efficient the method 
described in this paper could be for extracting the angle $\gamma$. 
With equal resonant and non-resonant
$\bbartocbar$ amplitudes, only the non-resonant regime ambiguities are
observed, due to the relative suppression of the resonant interference
terms discussed in Sec.~\ref{sec:res}.  Nonetheless, the $c_{\kstar
0}^\pm$ terms are significant enough to resolve all but the $\spi$
ambiguity. $\ssign$ is more strongly resolved, since it leaves neither
of the $c_{\kstar 0}^\pm$ terms invariant.

Also shown in Fig.~\ref{fig:scan-rnr} (dashed line) is the minimum
projection of $\chi^2(\gamma)$ of the average experiment. This plot is
obtained by averaging over $\chi^2(\gamma)$ of 50 simulated
experiments, each generated with the parameters of
Table~\ref{tab:reference}, but with different initial random numbers. The
three standard deviation region of $\gamma$ allowed by the average 
experiment is indicated. This region spans the range $[-0.06, 1.65]$,
giving an idea of the sensitivity that may be obtained with these parameter
values.

\begin{figure}[htb!]
\begin{center}
\epsfig{file=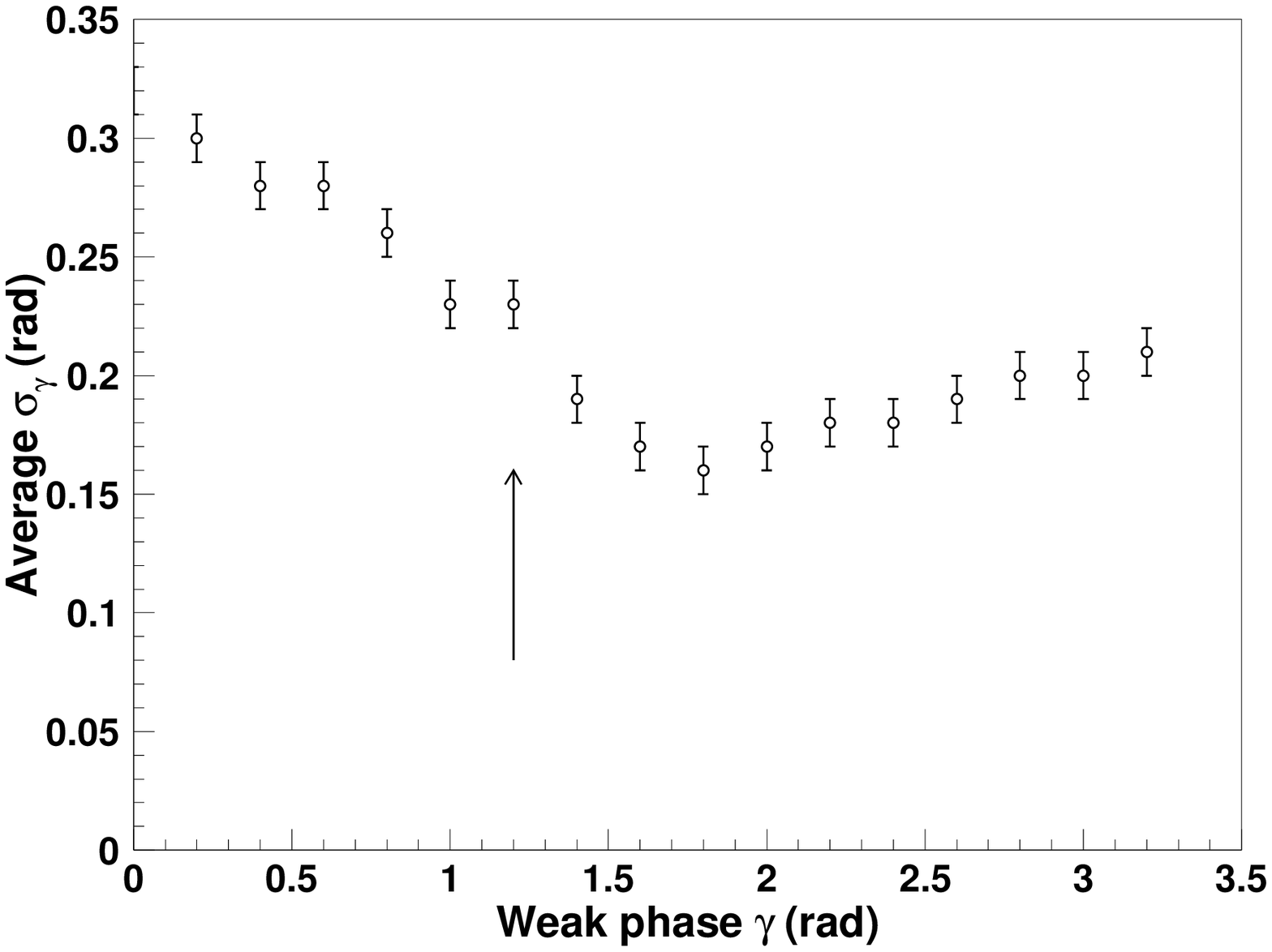,height=120mm} 
\caption{The error in $\gamma$, $\sigma_\gamma$, as a function of 
$\tilde\gamma$.}
\label{fig:sig-gamma-vs-gamma}
\end{center}
\end{figure}

\begin{figure}[htb!]
\begin{center}
\epsfig{file=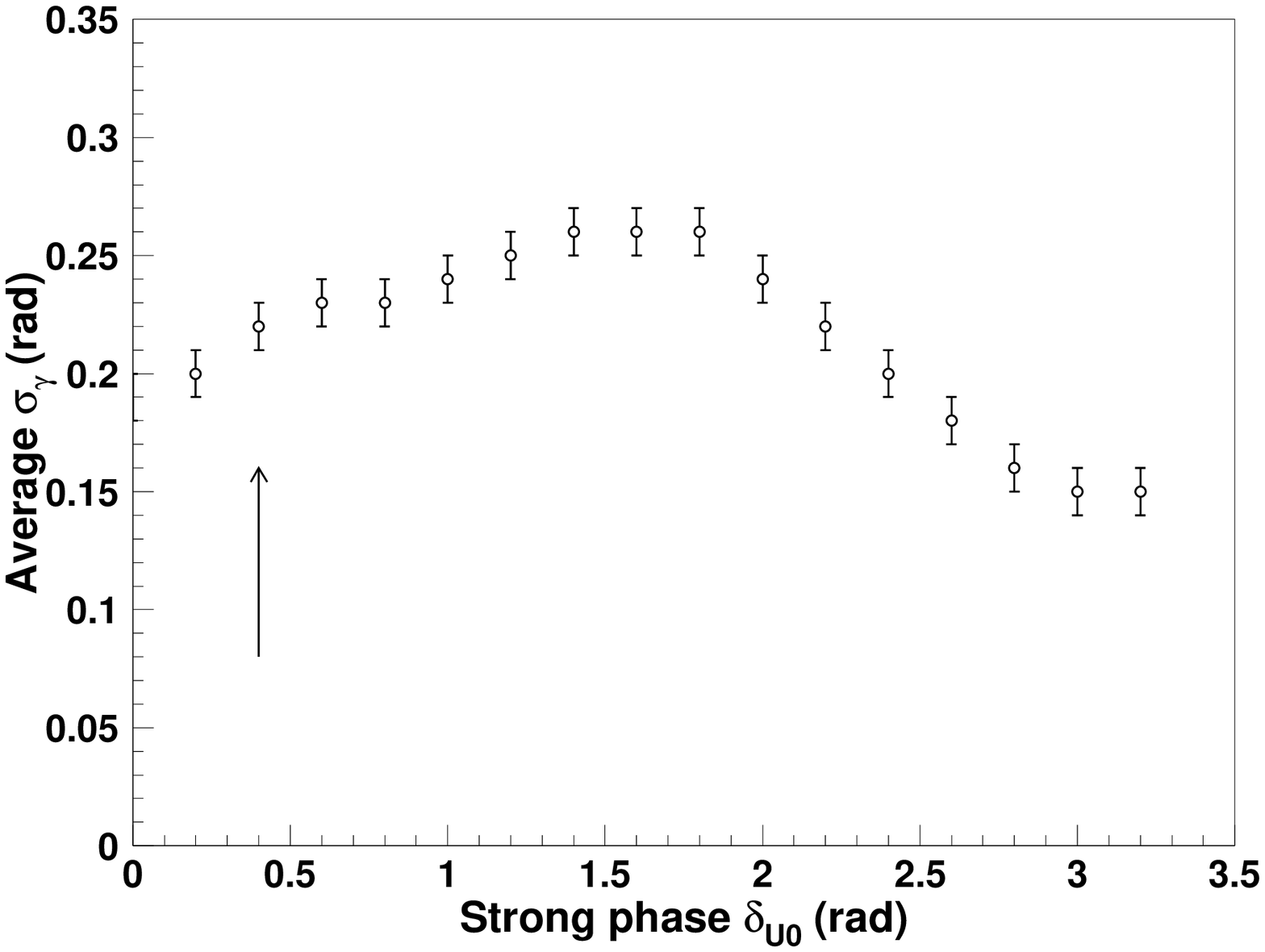,height=120mm} 
\caption{The error in $\gamma$, $\sigma_\gamma$, as a function of
$\tilde\delta\bu_0$.}
\label{fig:sig-gamma-vs-Delta}
\end{center}
\end{figure}

\begin{figure}[htb!]
\begin{center}
\epsfig{file=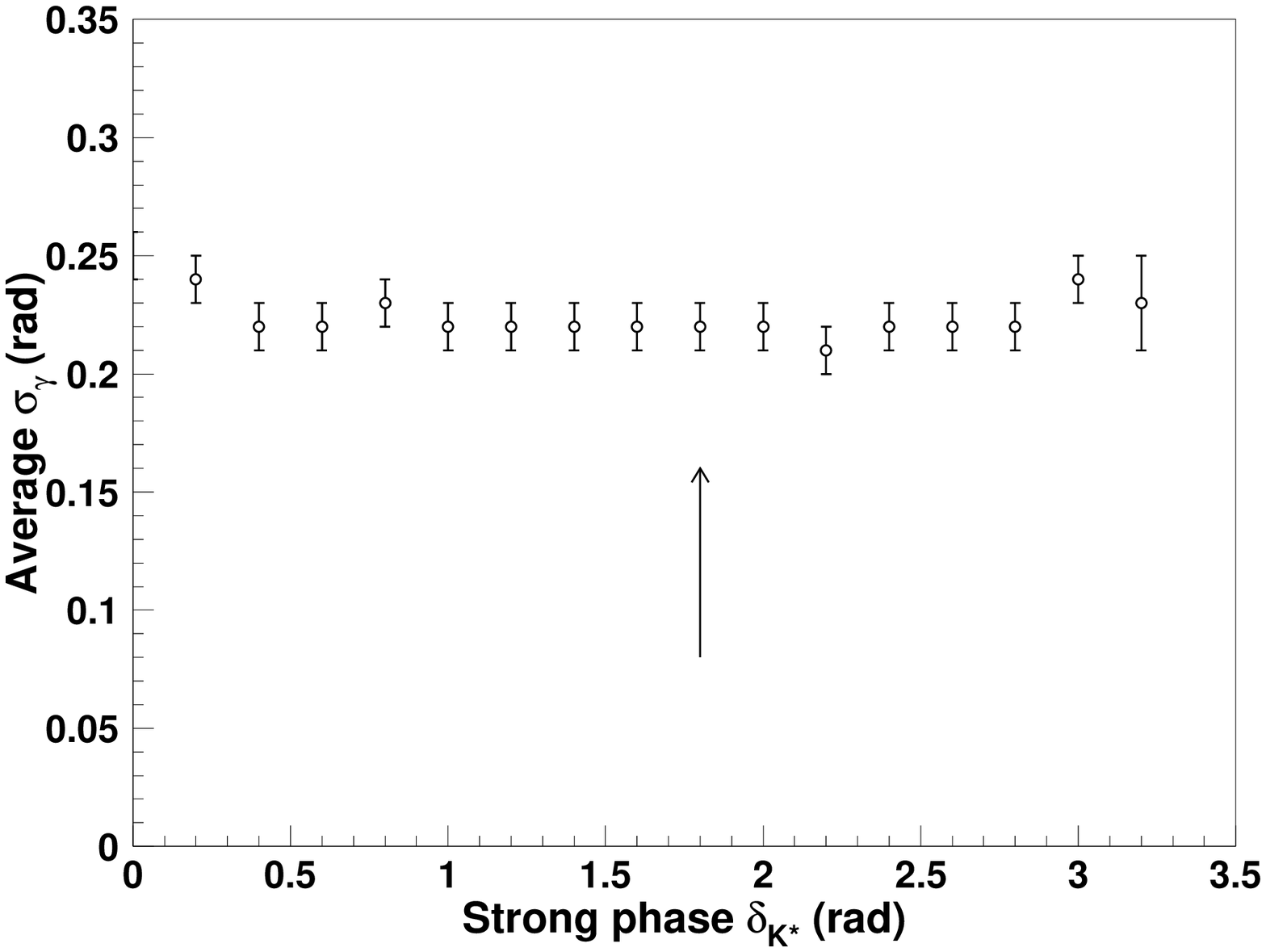,height=120mm} 
\caption{The error in $\gamma$, $\sigma_\gamma$, as a function of 
$\tilde\delta_{\kstar}$.}
\label{fig:sig-gamma-vs-delta-bc-kstar}
\end{center}
\end{figure}

\begin{figure}[htb!]
\begin{center}
\epsfig{file=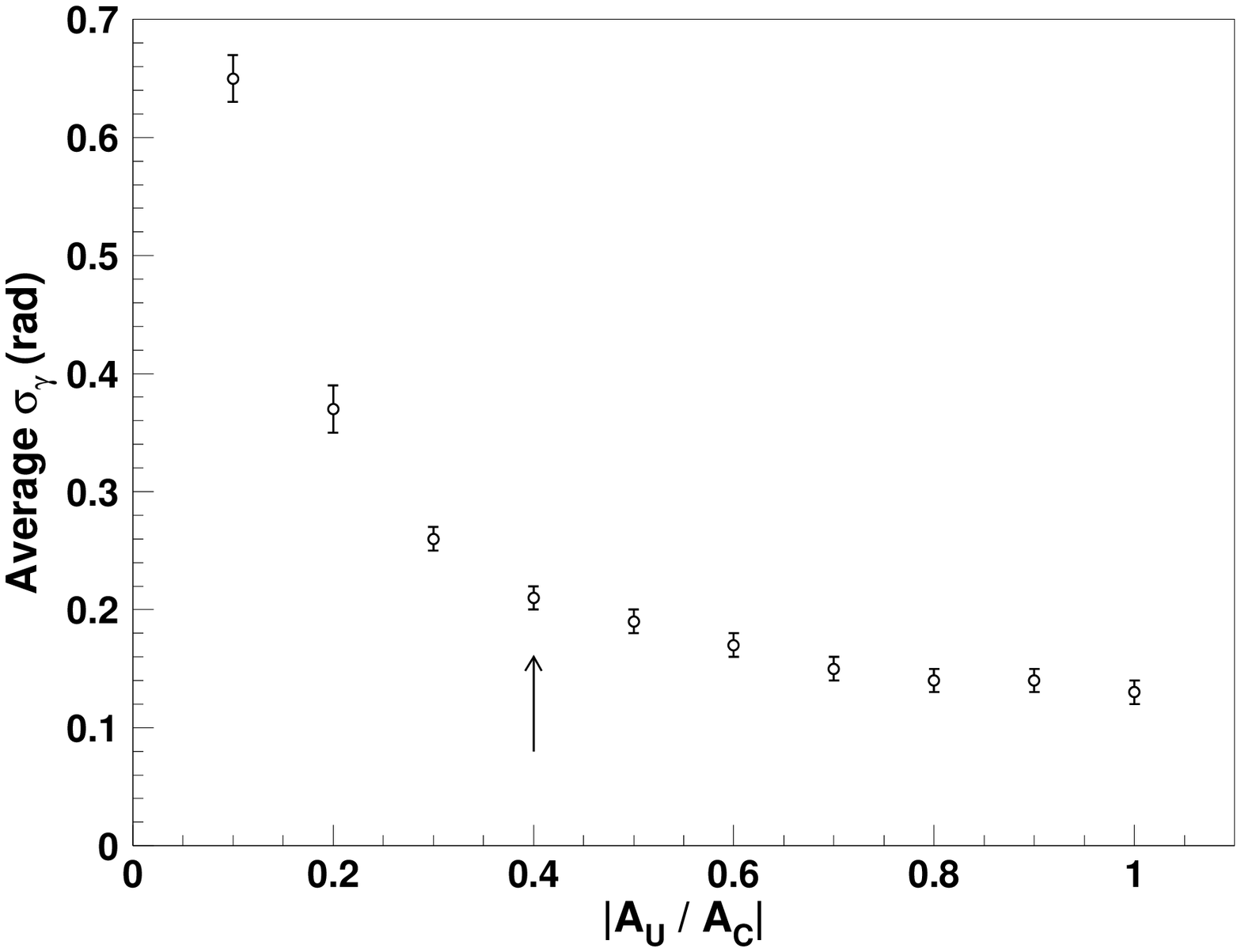,height=120mm} 
\caption{The error in $\gamma$, $\sigma_\gamma$, as a function of 
	$\tilde A\bu_0 / \tilde A\bc_0$.}
\label{fig:sig-gamma-vs-btou-over-btoc}
\end{center}
\end{figure}

In Figs.~\ref{fig:sig-gamma-vs-gamma}
through~\ref{fig:sig-gamma-vs-btou-over-btoc} we present
$\sigma_\gamma$, the statistical error in the measurement of $\gamma$,
obtained by fitting simulated event samples using the MINUIT
package~\cite{ref:minuit}, as a function of one of the parameters of
Table~\ref{tab:reference}.  All the other parameters were kept at the
values listed in Table~\ref{tab:reference}.
Each point in these plots is obtained by repeating the simulation 250
times, to minimize sample-to-sample statistical fluctuations. In all
cases, all the parameters of Table~\ref{tab:reference} were determined
by the fit.
The arrows in these figures indicate the
point corresponding to the parameters of Table~\ref{tab:reference}.
The total number of signal events in all final states combined
is the same for each of the data points.
The error bars describe the statistical error at each point, which is
determined by the number of experiments simulated.

One observes that $\sigma_\gamma$ does not depend strongly on
$\tilde\delta_{\kstar}$, and has a mild dependence on
$\tilde\delta\bu_0$. As expected, strong dependence on $\tilde A\bu_0
/ \tilde A\bc_0$ is seen in
Fig.~\ref{fig:sig-gamma-vs-btou-over-btoc}.  However, it should be
noted that $\sigma_\gamma$ changes very little for all values of
$\tilde A\bu_0 / \tilde A\bc_0$ above about 0.4, given that the total
number of signal events in all modes was kept constant in our
simulation. This suggests that the likelihood for a significantly
sensitive measurement is high over a broad range of parameters. With
the parameters of Table~\ref{tab:reference}, we find $\sigma_\gamma
\approx 0.23 = 13^\circ$ with an integrated luminosity of \lumi.

\section{Conclusions}
We have shown how $\gamma$ may be measured in the color-allowed decays
\btodkpiall, focusing on the simplest mode \btodkpmpiz.  The absence
of color suppression in the $\bbartoubar$ amplitudes is expected to
result in relatively large rates and significant CP violation effects,
and hence favorable experimental sensitivities. Although the Dalitz
plot analysis required for this purpose constitutes some experimental
complication, it should not pose a major difficulty, while being very
effective at reducing the eight-fold ambiguities that constitute a
serious limitation with other methods for measuring $\gamma$. Only the
two-fold $\spi$ ambiguity cannot be resolved solely by our method,
requiring additional constraints from other measurements of the
unitarity triangle.
As a result of these advantages, this method is likely to lead to
relatively favorable errors and provide a significant measurement of
$\gamma$, even with the current generation of B factory experiments.

\section{Acknowledgments}
The authors thank Francois Le Diberder for his fruitful ideas and help
with simulation.
This work was supported by the Danish Research Agency, the CEA and
CNRS-IN2P3 (France), and by the U.S. Department of Energy under
contracts DE-AC03-76SF00515 and DE-FG03-93ER40788.

\end{document}